\documentstyle[amssymb,aps,12pt,epsfig]{revtex}
\begin{document}
\title{Diffraction and trapping in circular lattices}
\author{H. L. Haroutyunyan and G. Nienhuis}
\address{Huygens Laboratorium, Universiteit Leiden,\\
Postbags 9504, \\
2300 RA Leiden, The Netherlands}
\maketitle

\begin{abstract}
When a single two-level atom interacts with a pair of Laguerre-Gaussian
beams with opposite helicity, this leads to an efficient exchange of angular
momentum between the light field and the atom. When the radial motion is
trapped by an additional potential, the wave function of a single localized
atom can be split into components that rotate in opposite direction. This
suggests a novel scheme for atom interferometry without mirror pulses. Also
atoms in this configuration can be bound into a circular lattice.
\end{abstract}


\section{Introduction}

It is well-known that light may carry both angular and linear momentum. When
a light field interacts with matter, exchange of momentum and angular
momentum between light and matter can occur. Laguerre-Gaussian (LG) light
modes are known to carry orbital angular momentum. If one employs the
paraxial approximation for the light field, simple expressions for the field
amplitudes and its average angular momentum can be derived \cite{Allen}. An
easy way to produce such beams is using spiral phase plates \cite{sumant}.

Another important question is the separability of the total angular momentum
into 'orbital' and 'spin' parts \cite{van Enk}. The orbital part is
associated with the phase distribution of the light field, and the spin part
is connected with its polarization. This question is essential in the
context of momentum transfer from light to the atom when one includes atomic
internal degrees of freedom. It has been shown that 'spin' and 'orbital'
angular momentum of the photon are transferred from the quantized light
field to, respectively, the internal and the external angular momentum of
the atom. The interaction with a LG mode is a possible way to entangle
internal and external degrees of freedom of an atom \cite{Muthu}. The
transfer of the angular momentum of light to particles has been also
experimentally demonstrated in \cite{Friese}, where trapped massive
particles are set into rotation while interacting with the light field.
Other authors have studied the cooling properties for atoms using LG beams
\cite{Kuppens}. Also LG beams have been proposed as a $2D$ trapping
potential for Bose condensates \cite{wright}.

Whereas angular momentum exchange between light and matter is a relatively
new topic, the linear momentum exchange is a well-established issue \cite
{Bernhardt}. It is well-known that two counterpropagating waves lead to a
more efficient exchange of linear momentum between an atom and the light
field than a single travelling wave. Using quantum language for a classical
light field, one can describe such an interaction as a sequence of
successive single photon absorption and emission events. This suggests that
one may expect more efficient angular momentum exchange between a light
field and an atom if one uses two LG modes with opposite helicity, e.g.
counterrotating waves.

\section{General framework}

We start with radiation fields propagating along the $z$-axis with wave
number $k$ and carrying orbital angular momentum (Laguerre-Gaussian beams).
If one considers the paraxial limit of these waves, the expressions for the
light fields are particularly simple \cite{Allen}
\begin{equation}
E\left( \rho ,z,\phi ,t\right) =E_{0}\left( \rho ,z\right) e^{i\left(
kz-\omega t+l\phi \right) }+c.c.,  \label{LG}
\end{equation}
where $\rho ,z,\phi $ are the cylindrical coordinates, $\omega $ is the
frequency and the integer index $l$ is indicates the helicity of the LG
beam. For two Laguerre-Gaussian beams with opposite helicity, namely $l$ and
$-l$, the total field can be written as
\begin{equation}
E\left( \rho ,z,\phi ,t\right) =2E_{0}\left( \rho ,z\right) \cos l\phi \text{
}e^{i\left( kz-\omega t\right) }+c.c.  \label{standing LG}
\end{equation}
We indicated already in the Introduction, that one expects a more efficient
exchange of angular momentum between the light field and the atom in the
configuration (\ref{standing LG}) than in a single LG mode. This expectation
is based on the corresponding situation of momentum exchange between an atom
and a standing light wave. In addition to the light field (\ref{standing LG}%
), the atomic motion in the radial direction is assumed to be confined by an
extra trapping potential $U\left( \rho \right) $ with cylindrical symmetry.

The $z-$dependence of the amplitude $E_{0}\left( \rho ,z\right) $ is slow
and can be ignored. Properly shaping the LG mode, the radial dependence of $%
E_{0}\left( \rho ,z\right) $ can be ignored on the characteristic width of
the trapping potential $U\left( \rho \right) $. Thus, we assume that $%
E_{0}\left( \rho ,z\right) \simeq E_{0}$ is constant. For a two-level atom
the Hamiltonian in the rotating-wave approximation can then be written as
\begin{equation}
\widehat{H}=\widehat{H}_{0}+U\left( \rho \right) +2\hbar \omega _{R}\cos
l\phi \left( e^{i\left( kz-\omega t\right) }\left| e\right\rangle
\left\langle g\right| +e^{-i\left( kz-\omega t\right) }\left| g\right\rangle
\left\langle e\right| \right) ,  \label{H}
\end{equation}
where $\omega _{R}$ is the Rabi frequency of each of the travelling waves
that create the standing wave, $\omega $ is the laser frequency, and
\begin{equation}
\widehat{H}_{0}=\frac{\widehat{P}^{2}}{2M}+\frac{\hbar \omega _{0}}{2}\left(
\left| e\right\rangle \left\langle e\right| -\left| g\right\rangle
\left\langle g\right| \right)  \label{H0}
\end{equation}
is the Hamiltonian for a free atom, with $\widehat{P}$ the momentum operator
of the atom, $\left| g\right\rangle $ and $\left| e\right\rangle $ indicate
the ground and excited states, and $\omega _{0}=(E_{e}-E_{g})/\hbar $
defines the transition frequency of a free atom.

The dynamics of the atom is rather simple if the laser is far detuned. We
assume that
\begin{equation}
\left| \Delta \right| \gg \omega _{R},  \label{Fardetuned}
\end{equation}
where the detuning $\Delta $ is defined as $\Delta =\omega _{0}-\omega .$
For an atom in the ground state, the excited state can be adiabatically
eliminated, which leads to an effective Hamiltonian in the well-known form
\begin{equation}
\widehat{H}=\frac{\widehat{P}^{2}}{2M}+U\left( \rho \right) +V\left( \phi
\right) ,  \label{Hamiltonian_far}
\end{equation}
where the light-shift potential is specified by
\begin{equation}
V\left( \phi \right) =-\hbar \Omega \cos ^{2}l\phi  \label{Trapping_pot}
\end{equation}
with $\Omega =%
%TCIMACRO{\dfrac{\omega _{R}^{2}}{\Delta }}
%BeginExpansion
{\displaystyle {\omega _{R}^{2} \over \Delta }}%
%EndExpansion
.$

\section{Trapping in counterrotating fields}

Some general conclusions on the bound states of the Hamiltonian (\ref
{Hamiltonian_far}) directly follow from its symmetry properties. We
introduce the unitary translation operator $\widehat{T}$ defined as
\begin{equation}
\widehat{T}\left| \phi \right\rangle =\left| \phi +\frac{\pi }{l}%
\right\rangle ,  \label{trans_operator}
\end{equation}
where $\left| \phi \right\rangle $ indicates the states with fixed azimuthal
angle. Since the Hamiltonian (\ref{Hamiltonian_far}) is invariant for
rotation about an angle $\pi /l$, it follows from a rotational version of
the Bloch theorem that the eigenstates of this Hamiltonian are also
eigenstates of $\widehat{T}$. The eigenvalue relation can be expressed as
\begin{equation}
\widehat{T}\left| \Psi _{q}\right\rangle _{j}=e^{-i\frac{\pi }{l}q}\left|
\Psi _{q}\right\rangle _{j},  \label{quasimomentum}
\end{equation}
where $q$ is referred to as angular quasimomentum and $j$ identifies the
energy band. We consider a single energy band, and we suppress the index $j$%
. We can restrict $q$ to the first Brillouin zone given as
\begin{equation}
-l\leq q<l.  \label{Brillouin}
\end{equation}
The eigenstates $\left| \Psi _{q}\right\rangle $ should be periodic in $\phi
$ with period $2\pi $, because a rotation over $2\pi $ must leave the wave
function invariant$.$ The finite range of $\phi $ leads to a discretization
of angular quasimomentum. On the other hand, a rotation over $2\pi $ is
equivalent to the action of the operator $\widehat{T}^{2l}$. Since it
follows from Eq. (\ref{quasimomentum}) that
\[
\widehat{T}^{\text{ }2l}\left| \Psi _{q}\right\rangle =e^{-2i\pi q}\left|
\Psi _{q}\right\rangle ,
\]
we conclude that the only possible values of the angular quasimomentum are
determined from the condition
\[
e^{-2i\pi q}=1.
\]
Hence $q$ must be integer, and each band contains $2l$ Bloch states. For
example, for $l=2$ the first Brillouin zone contains only the four values $%
q=-2,-1,0,1$ of the angular quasimomentum.

Also, in analogy to the case of an infinite linear lattice, one can
introduce localized Wannier states $\left| \Theta _{n}\right\rangle $ in the
usual manner, as Fourier transforms of the Bloch states
\[
\left| \Psi _{q}\right\rangle =\frac{1}{\sqrt{2l}}\sum_{n=-l}^{l-1}e^{iq%
\frac{\pi }{l}n}\left| \Theta _{n}\right\rangle .
\]
Obviously, the number of Wannier states within an energy band is equal to $%
2l $, just as the number of Bloch states.

In Fig. 1 we plot the trapping potential (\ref{Trapping_pot}) $V\left(
x,y\right) /\hbar \Omega $ for $l=2,4$ in Cartesian coordinates. When the
potential is sufficiently deep, atoms can be bound in the angular wells, and
the Wannier states are confined to a single well. An additional confining
potential $U(\rho )$ is required to trap particles in the radial direction,
and to avoid their escape. Then the potential (\ref{Trapping_pot}) can
create a circular lattice, where particles are located near the minima of
the periodic potential. A circular optical lattice has many applications, as
discussed recently by several authors \cite{circular}.

\section{Diffraction in counterrotating fields}

Since the potentials have a cylindrical symmetry, it is convenient to
express the kinetic energy in cylindrical coordinates, and we write
\begin{equation}
\frac{\widehat{P}^{2}}{2M}=-\frac{\hbar ^{2}}{2M}\left( \frac{\partial ^{2}}{%
\partial z^{2}}+\frac{1}{\rho }\frac{\partial }{\partial \rho }\rho \frac{%
\partial }{\partial \rho }+\frac{1}{\rho ^{2}}\frac{\partial ^{2}}{\partial
\phi ^{2}}\right) .  \label{P^2}
\end{equation}
The dynamics along the $z$ axis is completely free. For simplicity, we
assume that the radial potential $U(\rho )$ is narrow, so that the radial
motion is restricted to a ring with radius $\rho _{0}$, and we ignore radial
dispersion in the present Section. We return to it in Sec. VI, where the
effect of the radial dispersion is estimated. The motion of an atom in the
angular direction is then described by the one-dimensional Hamiltonian
\begin{equation}
\widehat{H}=-\frac{\hbar ^{2}}{2I}\frac{\partial ^{2}}{\partial ^{2}\phi }%
-\hbar \Omega \cos ^{2}l\phi ,  \label{Ham_azimuthal}
\end{equation}
which has the azimuthal angle as the only coordinate. The quantity $I=M\rho
_{0}^{2}$ is the moment of inertia. This Hamiltonian is the circular
counterpart of the Hamiltonian for simple linear diffraction. The main
difference is that the coordinate $\phi $ is periodic, which forces the
angular wave number $l$ to be integer. Diffraction of a single atom
described by such a linear Hamiltonian has been extensively studied
theoretically and experimentally by several groups \cite{Bernhardt}.

Just as is usually done for linear diffraction, we consider the situation
that an initially localized atom interacts with the optical potential during
a small interaction interval $[-\tau ,0]$, where the atom picks up momentum
from the lattice. The transition from the near field immediately after the
interaction and the far field is described by free evolution. We assume the
atom to be initially in its ground state and situated in a small segment of
the ring. Since the angular wave function $\Phi (\phi )$ of the atom must be
periodic at all times, we cannot represent a localized wave packet by a
Gaussian. The initial state at the beginning of the interaction interval is
taken as
\begin{equation}
\Phi \left( \phi ,-\tau \right) =C_{N}\cos ^{2N}\frac{\phi }{2},
\label{Psi(0)}
\end{equation}
with $N$ to be a large natural number, and $C_{N}$ is the normalization
constant
\begin{equation}
C_{N}=\frac{2^{2N}}{\sqrt{2\pi \left(
\begin{array}{l}
4N \\
2N
\end{array}
\right) }}.  \label{C_N}
\end{equation}
The state (\ref{Psi(0)}) can be written as a Fourier series, which is just
an expansion in the angular-momentum eigenstates. This gives
\begin{equation}
\Phi \left( \phi ,-\tau \right) =\frac{1}{\sqrt{2\pi }}\sum_{m=-N}^{N}\psi
_{m}e^{im\phi },  \label{ampl_momentum}
\end{equation}
with
\begin{equation}
\psi _{m}=\frac{1}{\sqrt{\left(
\begin{array}{l}
4N \\
2N
\end{array}
\right) }}\left(
\begin{array}{l}
2N \\
N+m
\end{array}
\right) .  \label{initial}
\end{equation}
The initial state (\ref{Psi(0)}) is localized around $\phi =0$,
which is clear from the asymptotic form
\begin{equation}
\cos ^{2N}\frac{\phi }{2}\simeq \exp \left\{ -\frac{N\text{ }\phi }{4}%
^{2}\right\} ,  \label{Gaus1}
\end{equation}
for large $N$. The half width in the azimuthal angle is of the order of $%
\sqrt{2/N}$. From the asymptotic form of the binomial coefficient
\[
\left(
\begin{array}{l}
2N \\
N+m
\end{array}
\right) \simeq 2^{2N}\frac{1}{\sqrt{\pi N}}\exp \left( -\frac{m^{2}}{N}%
\right)
\]
we find the asymptotic expression of the Fourier coefficient
\begin{equation}
\psi _{m}\simeq \left( \frac{2}{\pi N}\right) ^{1/4}\exp \left( -\frac{m^{2}%
}{N}\right) .  \label{initial_asympt}
\end{equation}
This demonstrates that the half width in angular momentum is of the order of
$\sqrt{N/2}$.

If we take the duration $\tau $ of the light pulse short and the moment of
inertia $I$ is large, so that $\hbar ^{2}l^{2}\tau /(2I)$, no propagation
occurs, and the kinetic-energy term can be neglected during the interaction.
This is the equivalence of the standard Raman-Nath approximation applied by
Cook et al \cite{Cook}. Then the final state at time $0$ after the
interaction is
\begin{equation}
\Phi \left( \phi ,0\right) =\Phi \left( \phi ,-\tau \right) \exp \left(
i\Omega \tau \cos ^{2}l\phi \right) .  \label{wave_function}
\end{equation}
This state can be expressed as an expansion in angular-momentum eigenstates,
in the form Fourier series, which is just an expansion in the
angular-momentum eigenstates. This gives
\begin{equation}
\Phi \left( \phi ,0\right) =\frac{1}{\sqrt{2\pi }}\sum_{m}\zeta
_{m}e^{im\phi },  \label{ampl_momentum0}
\end{equation}
where
\begin{equation}
\zeta _{m}=\exp \left( i\Omega t/2\right) \sum_{n}i^{n}\psi
_{m-2nl}J_{n}\left( \Omega \tau /2\right) ,  \label{Am}
\end{equation}
in terms of the ordinary Bessel functions.

States with large angular momentum $\left| m\right| >N$ are initially not
populated, whereas all angular momentum states get populated after the
interaction. Thus, the configuration with two LG modes leads to more
efficient exchange between the light field and the atom than a single LG
beam. The physical interpretation is the same as for diffraction in the
field of classical counterpropagating waves: an atom picks up a photon from
the light beam with one helicity and emits a photon into the opposite one.
In figure 2 we present a typical diffraction pattern calculated for the case
that $l>N.$ More precisely, we plot the angular-momentum coefficients $%
\left| \psi _{m}\right| ^{2}$ before the interaction, and the coefficients $%
\left| \zeta _{m}\right| ^{2}$ after the interaction with the circular
lattice, for $\Omega \tau =6.$ In the latter case, the momentum peaks
correspond to different values of $n$. The distance between neighboring
peaks is equal to $2l$. The half width of each peak is of the order of $%
\sqrt{N/2}.$

\section{Free evolution on a ring}

As shown above, the angular-momentum distribution of an atom after the
interaction with a pair of counterrotating LG beams can be broad. However,
as a result of the Raman-Nath approximation, the angular distribution of the
atom has not been modified during the interaction, so that $\left| \Phi
(\phi ,-\tau )\right| ^{2}=\left| \Phi (\phi ,0)\right| ^{2}$ . In this
chapter we investigate the spatial form of the atomic distribution in the
far field, i. e. after free evolution of the atom over the ring. As before,
the motion along the $z$ axis is completely free and the radial motion is
restricted on a ring. The initial state of this free evolution is given by
Eq. (\ref{wave_function}), with the expansion in angular-momentum states
given by Eqs. (\ref{ampl_momentum0}) and (\ref{Am}). For positive times, the
atomic motion is still restricted to the ring with radius $\rho _{0}$ by the
confining potential $U(\rho )$, and the evolution of the angular wave
function $\Phi (\phi )$ is governed by the Hamiltonian (\ref{Ham_azimuthal})
with $\Omega =0$. With the initial state (\ref{ampl_momentum0}), the
time-dependent wave function is given by the expansion
\begin{equation}
\Phi \left( \phi ,t\right) =\frac{1}{\sqrt{2\pi }}\sum_{m}\zeta _{m}\exp
\left( im\phi -i\xi tm^{2}\right)  \label{ampl_free}
\end{equation}
where $\xi =\hbar /(2I)$, and the coefficients $\zeta _{m}$ are given in Eq.
(\ref{Am}). As displayed in Fig. 2, the distribution $\left| \zeta
_{m}\right| ^{2}$ typically separates in a number of peaks centered at $%
\overline{m}=2nl$, where $n=0,\pm 1,$ $\pm 2,\ldots $which are separated by $%
2l$. Thus, the superposition state (\ref{ampl_free}) can be
considered as a series of elementary wave packets centered at
$2nl,$ in the angular-momentum space. Each of these peaks gives a
separate contribution to the wave
function that moves with its own angular group velocity $2\xi \overline{m}%
=4\xi nl\equiv \nu _{n}$. The angular separation between neighboring
wavepackets is given by $4\xi lt$, which is proportional to $l.$ Since wave
packets with opposite angular-momentum values will move in opposite
directions, i. e. clockwise and anticlockwise, they will eventually meet
again at some time $t=T$ and start to interfere.

In order to estimate the time value that interference sets in, we use the
fact that for not too small arguments $\Omega \tau /2$ the Bessel function $%
J_{n}(\Omega \tau /2)$ with the maximal value is the one with $n=n_{\max
}\simeq \Omega \tau /2$. Hence, the meeting time of the pair of the
strongest counterpropagating packets is
\[
T=\frac{\pi }{v_{\max }}=\frac{\pi }{2\xi \Omega \tau l}.
\]

The exact expression for the time-dependent wave function can be expressed
in an integral form by using the mathematical identity \cite{Berman}
\begin{equation}
\exp \left( im\phi -i\xi tm^{2}\right) =\frac{1}{\sqrt{4\pi i\xi t}}%
\int_{-\infty }^{\infty }d\phi ^{^{\prime }}\text{ }e^{im\phi ^{^{\prime
}}}\exp \left[ i\left( \phi -\phi ^{^{\prime }}\right) ^{2}/4\xi t\right] ,
\label{Math_rel}
\end{equation}
which can be checked by performing the integration. When substituting this
identity in the right-hand side of Eq. (\ref{ampl_free}), and using the
expansion (\ref{ampl_momentum0}), we arrive at the exact expression
\begin{equation}
\Phi \left( \phi ,t\right) =\frac{1}{\sqrt{4\pi i\xi t}}\int_{-\infty
}^{\infty }d\phi ^{^{\prime }}\text{ }\Phi (\phi ^{\prime },0)\exp \left[
i\left( \phi -\phi ^{^{\prime }}\right) ^{2}/4\xi t\right] .
\label{Psi_integr}
\end{equation}
A similar equation is well-known to describe the free evolution of a quantum
particle in one dimension. In the present case it is crucial that the
integration be performed over all values of $\phi ^{\prime }$, while using
that the wave function $\Phi (\phi ^{\prime },0)$ is periodic. Because of
this periodicity, we can express the integral in (\ref{Psi_integr}) as a sum
of bounded integrals

\begin{equation}
\Phi \left( \phi ,t\right) =\frac{1}{\sqrt{4i\pi \xi t}}\sum_{p=-\infty
}^{\infty }\int_{2\pi p}^{2\pi \left( p+1\right) }d\phi ^{^{\prime }}\text{ }%
\Phi (\phi ^{\prime },0)\exp \left[ i\left( \phi -\phi ^{^{\prime }}\right)
^{2}/4\xi t\right] .  \label{sum_2piL}
\end{equation}
By a shift of variables the integrations can be performed over the interval $%
[0,2\pi ]$, which leads to an integral expression over a single interval
\begin{equation}
\Phi \left( \phi ,t\right) =\frac{1}{\sqrt{4i\pi \xi t}}\sum_{p=-\infty
}^{\infty }\exp \left[ i\left( \phi -2\pi p\right) ^{2}/4\xi t\right]
\int_{0}^{2\pi }d\phi ^{^{\prime }}\text{ }\widetilde{\Phi }\left( \phi
^{^{\prime }},t\right) \exp \left[ -i\left( \phi -2\pi p\right) \phi
^{^{\prime }}/2\xi t\right] .  \label{sum_2piL1}
\end{equation}
Here we introduced the modified wave function $\widetilde{\Phi }$ which is
just the initial wave function, modified by a phase factor, defined by
\begin{equation}
\widetilde{\Phi }\left( \phi ^{^{\prime }},t\right) =\Phi (\phi ^{\prime
},0)\exp \left[ i\phi ^{^{\prime }\text{ }2}/4\xi t\right] .
\label{Psi_tilda}
\end{equation}
In order to emphasize its physical significance, we write Eq. (\ref
{sum_2piL1}) in the form
\begin{equation}
\Phi \left( \phi ,t\right) =\frac{1}{\sqrt{2i\xi t}}\sum_{p=-\infty
}^{\infty }\exp \left[ i\left( \phi -2\pi p\right) ^{2}/4\xi t\right]
F\left( \frac{\phi -2\pi p}{2\xi t}\right) ,  \label{sum_fourier}
\end{equation}
where the function $F$ is the Fourier transform of the modified wave
function defined over a single period
\begin{equation}
F\left( x\right) =\frac{1}{\sqrt{2\pi }}\int_{0}^{2\pi }d\phi ^{^{\prime }}%
\text{ }\widetilde{\Phi }\left( \phi ^{^{\prime }},t\right) \exp \left[
-ix\phi ^{^{\prime }}\right] .  \label{Fourier}
\end{equation}

For a freely evolving quantum particle in one dimension, the time-dependent
wave function has the same form as the term with $p=0$ in (\ref{sum_fourier}%
). The other terms can be understood from the periodic nature of the
dynamics on the circle, where each period of the initial wave function
serves as an additional source that contributes to the wave function $\Psi
\left( \phi ,t\right) $ in the relevant interval $[0,2\pi ]$. Because of the
finite range of the integration in Eq. (\ref{Fourier}), the distinction
between the modified wave function and the initial wave function vanishes
for times $t$ obeying the inequality $t\gg 1/(\xi N)$, when we find in a
good approximation
\begin{equation}
\widetilde{\Phi }\left( \phi ^{^{\prime }},t\right) \simeq \Phi \left( \phi
^{^{\prime }},0\right) .  \label{Zasympt}
\end{equation}
In this limit, the function $F$ is just the Fourier transform of the initial
wave function $\Phi (\phi ,0)$, and $\Phi (\phi ,t)$ is simply determined by
the Fourier transform $F$ of the initial wave function $\Phi (\phi ,0)$
multiplied by a phase factor. The equation (\ref{sum_fourier}) has the
flavor of the far-field picture of the time-dependent wave function. The
Fourier transform of the initial state determines not only the momentum wave
function, but also the asymptotic form of the coordinate wave function,
scaled by a factor that varies linearly with time. Characteristic for the
present case of evolution on a circle is that each interval of length $2\pi $
serves as a separate source, each giving a contribution to $\Phi (\phi ,t)$.
Since the Fourier transform of the wave function determines the
angular-momentum amplitudes, we may conclude that the wave function for not
too small times has the same form as the initial distribution of angular
momentum, scaled by the factor $2\xi t$.

It is clarifying to follow the temporal evolution of $\left| \Phi \left(
\phi ,t\right) \right| ^{2}$ by distinguishing two time regions, namely $%
0\leq t<T$ and $t\geq T.$ In the region $0\leq t<T$, the wave function has
not yet spread beyond a single period of length $2\pi $, and only a single
term in eq. (\ref{sum_fourier}) (or (\ref{sum_2piL1})) differs from zero.
The contribution to the wave function coming from different sources do not
overlap yet, so that one can neglect the interference term between them. At
later times $t\geq T,$ the diffraction pattern on the interval $\left[
0,2\pi \right] $ is formed as an interference pattern between two and more
terms in the superposition state (\ref{sum_2piL}). This picture is confirmed
by numerical calculation of the diffraction pattern for the two time
regimes. In Fig. 3 the angular probability distribution $\left| \Phi \left(
\phi ,t\right) \right| ^{2}$ is shown for a time $t<T$. The spatial pattern
resembles the angular momentum distribution shown in Fig. 2. Figure 4
displays the same probability distribution for a later time $t>T$. One
notices that the counterrotating components give rise to clear interference
fringes. These fringes will be quite sensitive to any perturbation in one of
the arms. This suggests to use the present scheme as an atomic
interferometer \cite{Berman}. Usually, interferometers have two key
components, namely a beam splitter and a mirror. A coherent incoming atomic
beam is split into spatially separated components by the beam splitter. Two
arms are getting formed, which freely propagate and may undergo different
phase shifts, which are probed by recombining the two arms. The interference
pattern contains the information of the phase perturbation in one of the
arms. Recombination usually requires atomic mirrors. In atom optics, beam
splitters and mirrors are commonly realized by using light pulses, with
carefully selected duration and shape.

In the present case, only a single pulse is required that splits the initial
atomic wave packet into components rotating to the left and to the right. No
mirrors are employed in this scheme. Instead, one uses the radial potential $%
U\left( \rho \right) $, to constrain the atomic motion to a ring. Radial
potentials can be realized by hollow light beams, which are widely used in
atomic interferometric schemes.

\section{Radial dispersion}

In this chapter we consider the radial dynamics of the diffracted wave
function during its free evolution, after the passage of the circular
lattice. We assume that the wave function at time $t=0$, after the
diffracting pulse, is factorized as
\begin{equation}
\Psi (\rho ,\phi ,0)=Q(\rho ,0)\Phi (\phi ,0),  \label{initial2}
\end{equation}
where the radial part $Q$ of the wave function is sharply peaked at $\rho
=\rho _{0}$, and the angular wave function is specified by Eq. (\ref
{ampl_momentum0}). The radial function $Q$ is normalized ($\int_{0}^{\infty
}d\rho Q^{2}(\rho )\rho =1$). We wish to study the possible deformation of
the wave packet, when the radial dispersion is included during the stage of
free evolution. We take the simplest possible trapping potential, which
allows radial dispersion, and we take for the confining potential $U\left(
\rho \right) $ an infinitely deep cylindrical box with radius $a$, as
defined by
\begin{equation}
U\left( \rho \right) =0\text{ for }\rho \leq a,\text{ and }U\left( \rho
\right) =\infty \text{ for }\rho >a.  \label{Hollow}
\end{equation}
This potential models a hollow light beam. With this potential, the
normalized eigenfunctions of the Hamiltonian for the cylindrical coordinates
during the free-evolution stage take the form
\begin{equation}
\Psi _{nm}(\rho ,\phi )=R_{nm}(\rho )\frac{1}{\sqrt{2\pi }}e^{im\phi },
\label{eigencyl}
\end{equation}
where the radial functions $R_{nm}$ are solutions of the equation
\begin{equation}
\left[ -\frac{\hbar ^{2}}{2M}\left( \frac{1}{\rho }\frac{\partial }{\partial
\rho }\rho \frac{\partial }{\partial \rho }-\frac{m^{2}}{\rho ^{2}}\right)
+U\left( \rho \right) \right] R_{nm}(\rho )={\cal E}_{nm}R_{nm}(\rho ),
\label{Ham_free_eq}
\end{equation}
with ${\cal E}_{nm}$ are the corresponding eigenenergies. The radial
functions are proportional to the Bessel function of order $m$
\begin{equation}
R_{nm}(\rho )\varpropto J_{m}(\alpha _{nm}\rho /a),  \label{eigenfunction}
\end{equation}
with $R_{nm}$ normalized in the interval $0\leq \rho \leq a$. In order that
the wave function vanishes at the edge $\rho =a$ of the cylindrical well, we
have to take the numbers $\alpha _{nm}$ for various values of $n$ as the
subsequent zero's of the Bessel function $J_{m}$. This determines the
corresponding eigenenergies as

\begin{equation}
{\cal E}_{nm}=\hbar \lambda \alpha _{nm}^{2}.  \label{energy}
\end{equation}
with $\lambda =\hbar /(2Ma^{2})$. For each value of the angular momentum $m$%
, the set of functions $R_{nm}(\rho )$ is complete. An expansion of the
initial state (\ref{initial2}) in the energy eigenfunction is found when we
expand the initial radial wave function $Q(\rho ,0)$ in the radial
eigenfunctions (\ref{eigenfunction}), so that
\begin{equation}
Q(\rho ,0)=\sum_{n}c_{nm}R_{nm}(\rho ),  \label{expans0}
\end{equation}
while substituting Eq. (\ref{ampl_momentum0}) for the initial angular state $%
\Phi (\phi ,0)$. For the time-dependent state we find
\begin{equation}
\Psi (\rho ,\phi ,t)=\sum_{m}\frac{1}{\sqrt{2\pi }}\zeta _{m}e^{im\phi
}Q_{m}(\rho ,t),  \label{totalt}
\end{equation}
where the $m$-dependent radial wave function $Q_{m}$ is
\begin{equation}
Q_{m}(\rho ,t)=\sum_{n}c_{nm}R_{nm}(\rho )\exp (-i{\cal E}_{nm}t/\hbar ).
\label{radialt}
\end{equation}
>From Eq. (\ref{expans0}) one notices that $Q_{m}(\rho ,t)=Q(\rho ,t)$,
independent of the angular momentum $m$. It is obvious from the radial
Schr\={o}dinger equation (\ref{Ham_free_eq}) and the initial condition (\ref
{initial2}) that the normalized radial wave function obeys the identity $%
Q_{m}(\rho ,t)=Q_{-m}(\rho ,t)$ for all $m$. Moreover, since the total wave
function before diffraction is even in $\phi $, it must remain even for all
times. This implies that $\zeta _{m}=\zeta _{-m}$ for all $m$. So just as
discussed in Sec. IV, the angular distribution separates in different wave
packets that are counterrotating. Since the phase of $\zeta _{m}Q_{m}$ is
even in $m$, its derivative with respect to $m$ will be odd, and the angular
group velocities of packets with opposite values of $\overline{m}$ will be
opposite. This leads to interference after the packets have traversed the
entire ring. The initial radial function is taken as a narrow Gaussian
\begin{equation}
Q\left( \rho ,0\right) \propto \exp \left( -\left( \rho -\rho _{0}\right)
^{2}/2L^{2}\right) ,  \label{Psi_r_initial}
\end{equation}
Here $L$ is the width and $\rho _{0\text{ }}$represents the initial position
of the wave packet within the box. The normalized wave function $Q_{m}(\rho
,t)$ describes the radial dynamics for each value of the angular momentum $m$%
. As an example, we evaluate the time behavior of the average radius for
each angular momentum, with the given initial radial state (\ref
{Psi_r_initial}), according to the expression

\[
\left\langle \rho \left( t\right) \right\rangle _{m}=\int_{0}^{\infty }d\rho
\left| Q_{m}(\rho ,t)\right| ^{2}\rho ^{2}.
\]
The result is displayed in Fig. 5, in the special case that
$m=10$. The average radius displays oscillations, which can be
understood as arising from the outward motion due to the
centrifugal potential, followed by reflection at the hard wall of
the cylinder. The oscillations display collapse, followed by a
revival. These may be viewed as arising from the initial dephasing
of the contribution from the radial eigenfunctions $R_{nm}$ with
different values of $n$, due to their energy difference. The
revival of the oscillation can be understood from the discrete
nature of the contributing energy eigenvalues, when the phase
factors due to neighboring eigenenergies have built up a phase
difference $2\pi $. Because of the conservation of angular
momentum, the probability density near the origin remains zero.
The interference between the counterrotating wave packets is
illustrated in Fig. 6, for the ring at radius $\rho =\rho
_{0}=a/2$. Fig. 6a shows the short-time separation of the angular
wave packets. Fig. 6b displays the interference that arises as
soon as overlap occurs around $\phi =\pi $ between the clockwise
and the anti-clockwise rotating packets. This demonstrates that
the radial wave functions $Q_{m}(\rho ,t)$ for different values of
$m$ have sufficient overlap, so that the angular interference
survives the effect of radial dispersion.

\section{Conclusions}

In this paper we describe the diffraction of an atomic wave by a circular
optical lattice. Such a lattice can be formed by the superposition of two
Laguerre-Gaussian beams with opposite helicity, which gives rise to a
standing wave in the angular direction. Such a light field will split a
single localized atom into clockwise and anticlockwise rotating components.
If the system is in a trapping potential in the form of a ring or in a
cylindrical box, these counterrotating components give rise to interference.
We express the spatial pattern in the far diffraction field in terms of the
Fourier transform of the near-field diffraction pattern. The periodic nature
of the circular motion modifies this relation compared with the case of
diffraction by a linear standing wave. The general conclusions are backed up
by numerical calculations. Characteristic for the circular case is that the
wave packets corresponding to opposite angular momentum will cross each
other, even without applying light pulses to reverse their motion, as in
more common interferometric schemes. The scheme is reasonably robust to
changes in the radial confining potential.

\acknowledgments
This work is part of the research program of the ``Stichting voor
Fundamenteel Onderzoek der Materie'' (FOM).

\newpage
\begin{figure}[tbh]
\centerline{\psfig{figure=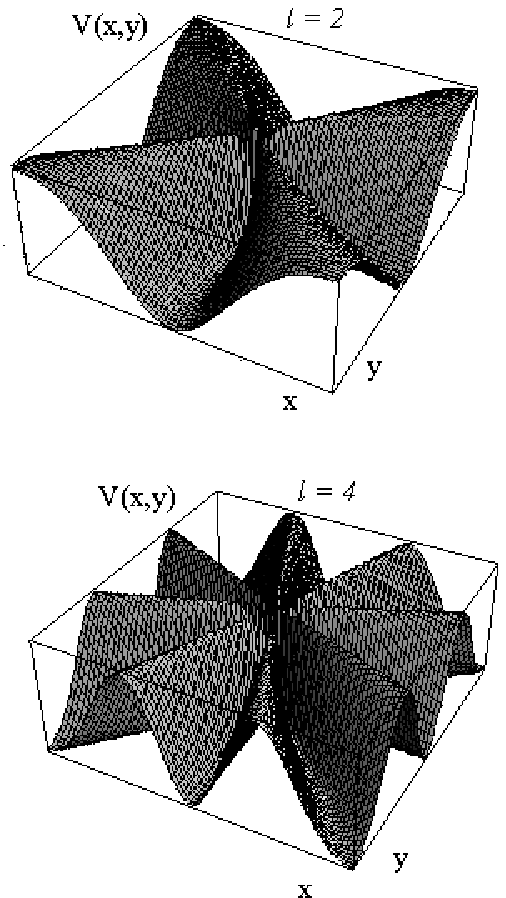,width=6cm}}
\caption{Circular
lattice structure due to the trapping potential $V\left(
x,y\right) $. The plot shows $V\left( x,y\right) /\hbar \Omega $
for $l=2, 4 $ in Cartesian coordinates.}
\end{figure}
\newpage

\begin{figure}[tbh]
\centerline{\psfig{figure=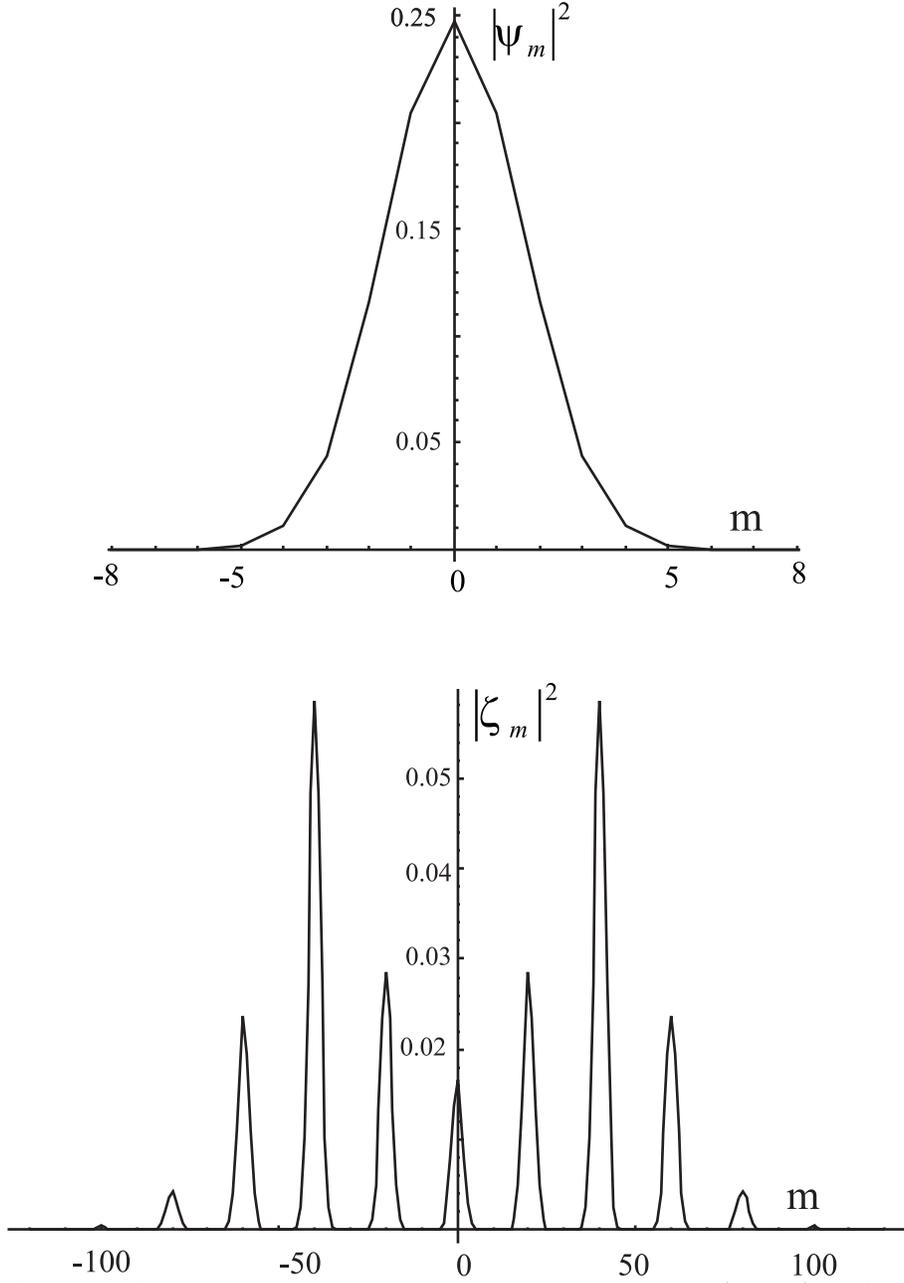,width=12cm}}
\caption{Probability distribution of angular momentum $m$ before
(upper) and after the pulse (lower). Here the helicity of the circular lattice is $l=10$%
, the initial state is determined by $N=10$ and the pulse duration $\tau$ is
given by $\Omega \tau=6. $}
\end{figure}
\newpage

\begin{figure}[tbh]
\centerline{\psfig{figure=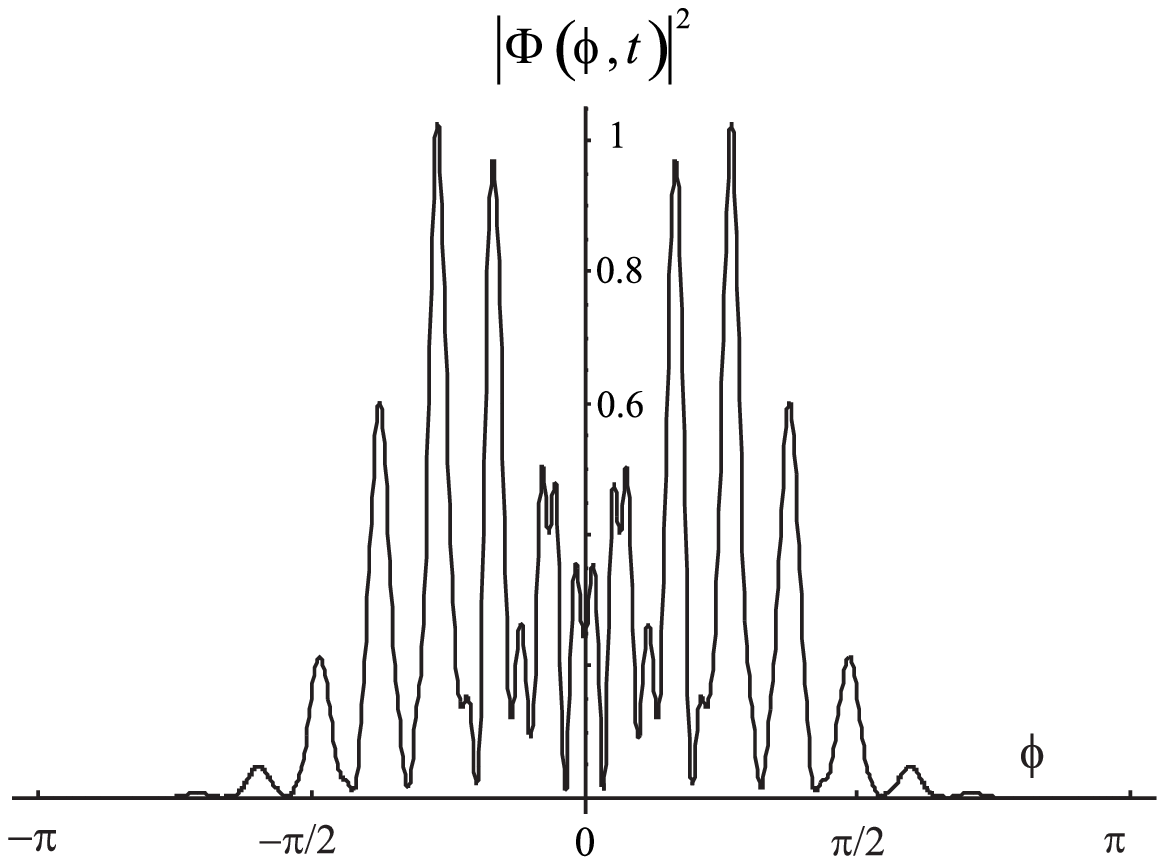,width=12cm}}
\caption{Angular
distribution $\left|\Phi \left( \phi ,t\right) \right| ^2$ is
plotted versus the azimuthal angle $\phi $ before the left and
right rotating components cross. Here $\xi t=3\pi \times 10^{-3}$,
the value of $N$ determining the width of the initial state, the
helicity $l$ and the pulse duration $\tau$ are the same as in Fig.
2.}
\end{figure}
\newpage

\begin{figure}[tbh]
\centerline{\psfig{figure=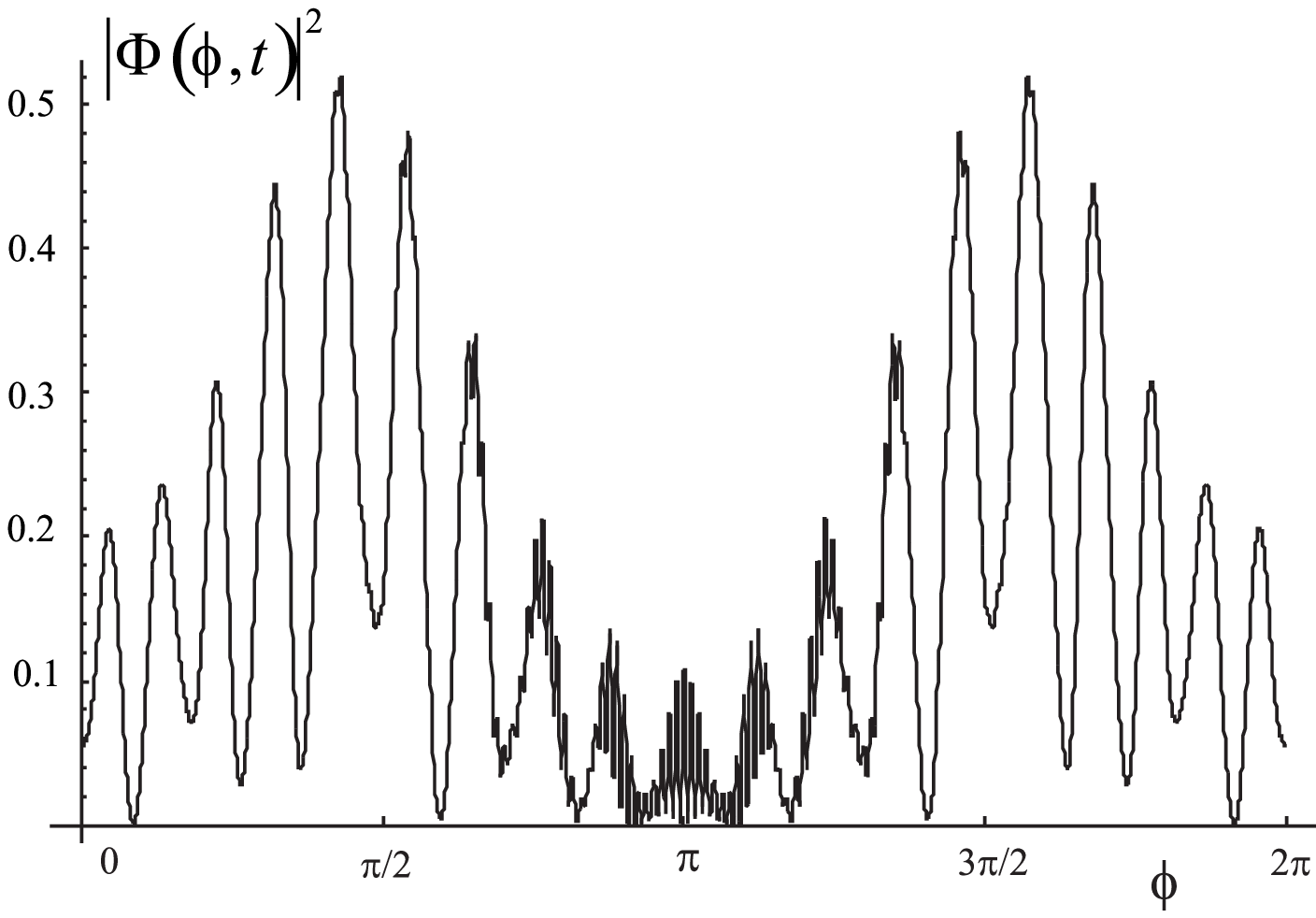,width=12cm}}
\caption{Angular
distribution $\left|\Phi \left( \phi ,t\right) \right| ^2$ is
plotted versus the azimuthal angle $\phi $ after the left and
right
rotating components cross. Here $\xi t=6\pi \times 10^{-3}$, the value of $N$%
, the helicity $l$ and the pulse duration $\tau$ are the same as
in Fig. 2.}
\end{figure}
\newpage

\begin{figure}[tbh]
\centerline{\psfig{figure=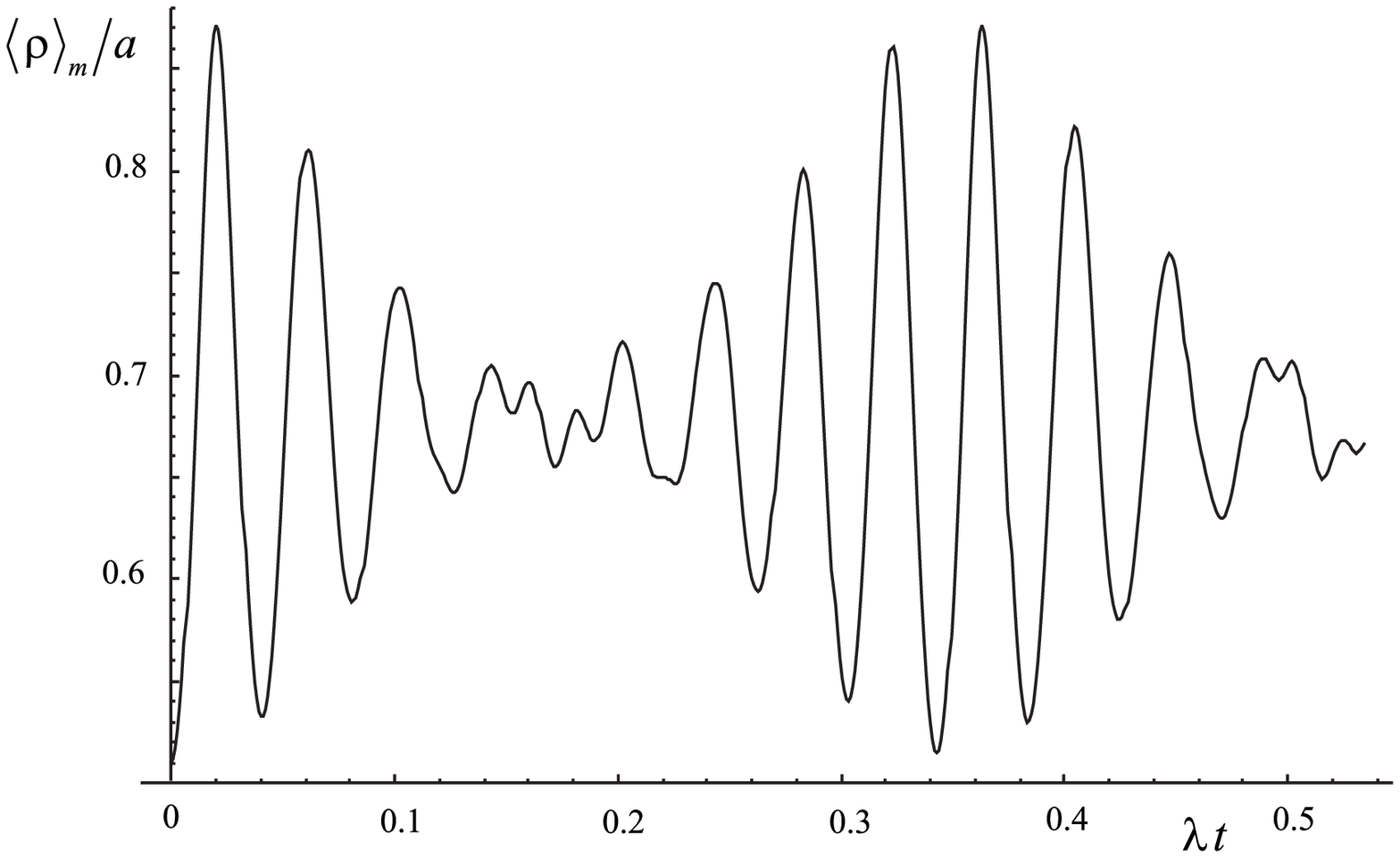,width=12cm}}
\caption{Time
behavior of the average radial distance $\left\langle \rho \left(
t\right) \right\rangle_m /a$ for the angular momentum $m=10.$ $\
N$ is
the same as in Fig. 2, the width of the initial Gaussian is $%
L=0.01a$, and the initial average radial distance is
$\rho_0=a/2$.}
\end{figure}
\newpage

\begin{figure}[tbh]
\centerline{\psfig{figure=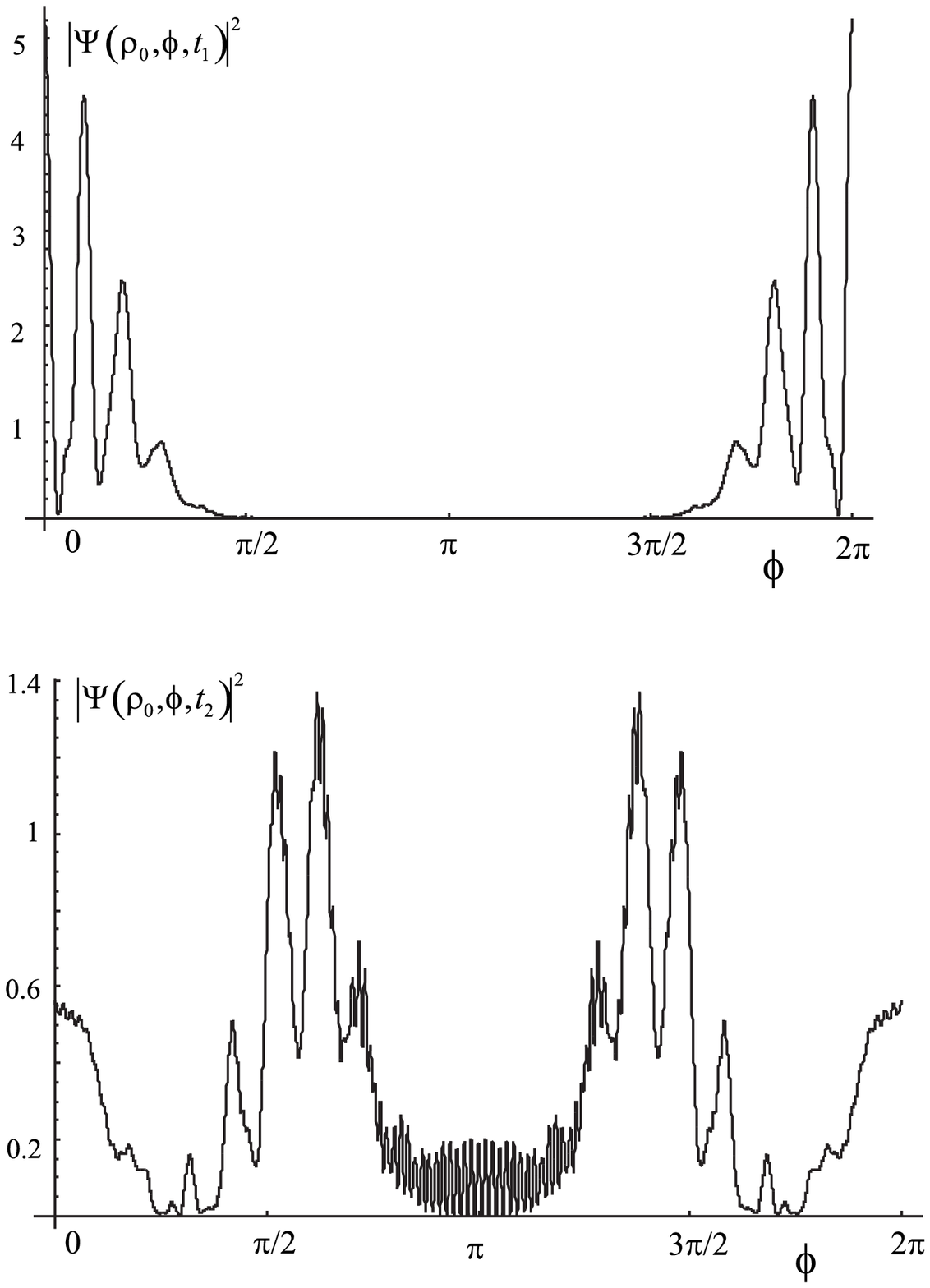,width=12cm}} \caption{Angular
distribution $\left|\Psi \left(\rho_0, \phi ,t\right) \right| ^2$
in the presence of radial dispersion, at the ring $\rho = \rho_0 =
a/2.$ The time values are determined by $\lambda t_{1} = \pi
\times 10^{-3} $ and $\lambda t_{2} = 2\pi \times 10^{-3}$. $N, l,
\Omega \tau$ are the same as in Fig. 2. }
\end{figure}
\newpage

\end{document}